\documentstyle[sprocl,epsfig]{article}

\def\Journal#1#2#3#4{{#1} {\bf #2}, #3 (#4)}

\def\NPA{{\em Nucl. Phys.} A}

\def\PRC{{\em Phys. Rev.} C}

\newfam\BMath
\font\BMathL=cmmib10 
\font\BMathl=cmmib7
\font\BMathm=cmmib5
\textfont\BMath=\BMathL \scriptfont\BMath=\BMathl
\scriptscriptfont\BMath=\BMathm

\def\mthbf#1{{\fam\BMath #1}}

\def\a{\alpha}
\def\b{\beta}
\def\d{\delta}
\def\g{\gamma}
\def\o{\omega}
\def\p{\pi}
\def\t{\tau}

\def\cm{{\cal M}}

\def\K{{\mathbf k}}

\def\pd{\partial}

\def\lra{\longrightarrow}
\def\llra{\longleftrightarrow}

\def\nonum{\nonumber \\}
\def\fx{\!}
\def\intkso{\frac{d^3 \K_1}{(2\pi)^3 2\o_1}}
\def\intkst{\frac{d^3 \K_2}{(2\pi)^3 2\o_2}}
\def\intksth{\frac{d^3 \K_3}{(2\pi)^3 2\o_3}}

\def\be{\begin{equation}}
\def\ee{\end{equation}}
\def\bea{\begin{eqnarray}}
\def\eea{\end{eqnarray}}
\def\bfig{\begin{figure}}
\def\efig{\end{figure}}
\def\eref#1{Eq.~(\ref{#1})} 
\def\fref#1{Fig.~\ref{#1}}

\font\bbf=cmb7

\begin{document}

\title{EQUILIBRATION AND OUT-OF-EQUILIBRIUM EFFECT IN RELATIVISTIC
HEAVY ION COLLISIONS}

\author{S.M.H. WONG\footnote{Present address: School of Physics and 
Astronomy, University of Minnesota, Minneapolis, 
MN 55455, U.S.A.}}

\address{Nuclear and Particle Physics Section, University of Athens,\\
Panepistimiopolis, GR-15771 Athens, Greece}
\address{Institute of Accelerating Systems and Applications (IASA),\\
P.O. Box 17214, GR-10024 Athens, Greece}
\address{and\\ Fachbereich Physik, Universit\"at Wuppertal, 
D-42097 Wuppertal, Germany}


\maketitle

\abstracts{The approach of a parton plasma at future
heavy ion colliders towards kinetic and chemical equilibrium 
is considered. A plasma with a self-consistent evolving 
parton-parton interaction strength is shown to equilibrate 
better and faster than the usual but inconsistent one with a 
fixed strength. We explain why as a consequence of this, a 
parton plasma is a unique kind of many-body system. Because our 
time evolution scheme does not require the plasma to be in either 
kind of equilibrium from the outset, out-of-equilibrium effect on 
particle productions can be revealed. We show this on photon 
production and discuss the implications on photon as a signal to 
detect the quark-gluon plasma.
\hfill{\bbf IASA 98-3, UA/NPPS-98-13}
}
\section{Introduction}

The trophy of the game of relativistic heavy ion collisions
at the future Relativistic Heavy Ion Collider (RHIC) and the
Large Hadron Collider (LHC) is well known, namely the
quark-gluon plasma. To reach this goal, one interrogates
the particles coming out from the beginning till the
very end of the collisions. Based on the information thus
obtained, one determines whether a phase transition into
the predicted nearly free quarks and gluons state has occurred.
In order to perform the necessary analysis, qualitative understanding
and quantitative control are very important. In this talk,
we present some recent development in the former.
We will look at the equilibration of the parton plasma in 
the following sections to determine the state of the system
just before the phase transition. One of the reasons being
that if the system is able to complete the equilibration
process, then it would be much simpler to describe.
The formalism of thermal field theory for QCD would then be
applicable. If the system is, on the other hand, still out
of equilibrium, this much more complicated situation would
require a phenomenological approach as the out-of-equilibrium
thermal field theory is not yet fully developed. Another reason 
is the state of equilibrium of the system has important effects on 
the particle emissions or their production rates. We will show 
this effect on photon production from the plasma. Before
we do that, the parton plasma as a rather unique many-body
system will be shown first.

\section{Equilibration of a parton plasma --- fixed vs. 
evolving $\a_s$}

To study the equilibration of the parton plasma, we must have 
a means of performing the time evolution of such a many-body
system. To do this, we choose some plausible initial conditions
from event generator such as HIJING. Then we build the time
evolution equations from Boltzmann equation
\be \left ( {\pd \over {\pd t}} 
           +{\mthbf v}_p \cdot {\pd \over {\pd \mthbf r}} \right )
    f(p,\t) = C(p,\t)  \; .
\ee
The collision terms $C(p,\t)$ will be constructed in two ways
in order to close the equations. The first is to use the
relaxation time approximation and the second is by explicit
construction from QCD matrix elements and the particle
distributions $f_g$, $f_q$ and $f_{\bar q}$. We include 
all binary interactions at the leading order in the 
renormalized strong coupling
\bea  gg \llra gg  \;\;\; & \;\;\; qg \llra qg \;\;\; & \;\;\; 
      \bar q g \llra \bar q g  \nonum
      qq \llra qq  \;\;\; & \;\;\; q\bar q \llra q \bar q \;\;\; & \;\;\; 
      \bar q \bar q \llra \bar q \bar q                                
\eea
and in order to have non-conservation of the number of partons,
which one would expect from interactions generated by the
non-Abelian QCD Lagrangian, we include also
\be  gg \llra ggg \mbox{\hskip 2.0cm} gg \llra q\bar q  \; .
\ee
Combining all these ingredients, we can solve for the distributions,
which are functions of time and from which most information
of the plasma can be obtained and therefore the time evolution
is known. Of course, it is still necessary to choose a most
appropriate value of the coupling to evaluate the collision
terms. For this purpose, it is customary to assume an average
momentum transfer of 2 GeV for the parton interactions in the 
plasma. This translates into an $\a_s=0.3$ for $\Lambda_{QCD} = 200$ 
MeV. With this choice, the system can be evolved in time and one can 
check for the state of equilibrium. Concerning the latter, there 
are two aspects. One is the balance of the partonic composition
in the plasma or parton chemical equilibration. The degree
of chemical equilibrium is most simply parametrized by the
quantity known as the fugacity which we denote here by 
$l_g$ and $l_q$ for gluons and quarks respectively. 
The plasma is in chemical equilibrium if $l=1.0$ or very
far from equilibrium if $l=0.0$.
In \fref{f:fug}, the results of the change of the fugacities
with time at LHC and at RHIC are shown in solid lines.
As can be seen, chemical equilibration for gluon is very good
at LHC and good at RHIC. On the contrary, that for quarks
at both colliders is not so good and even poor at RHIC.

\bfig
\centerline{
\hbox{
\epsfig{figure=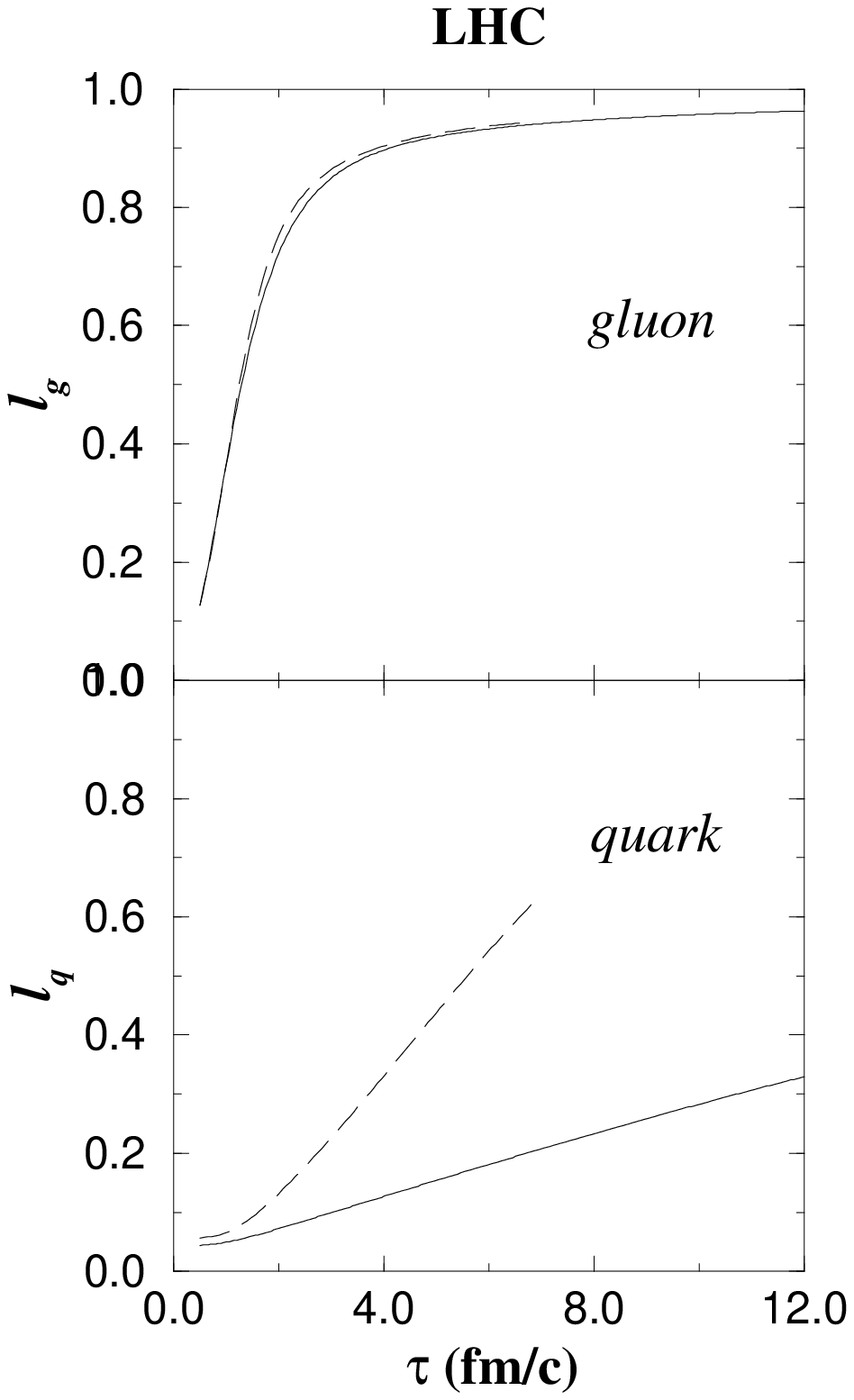,width=1.60in} \ \
\epsfig{figure=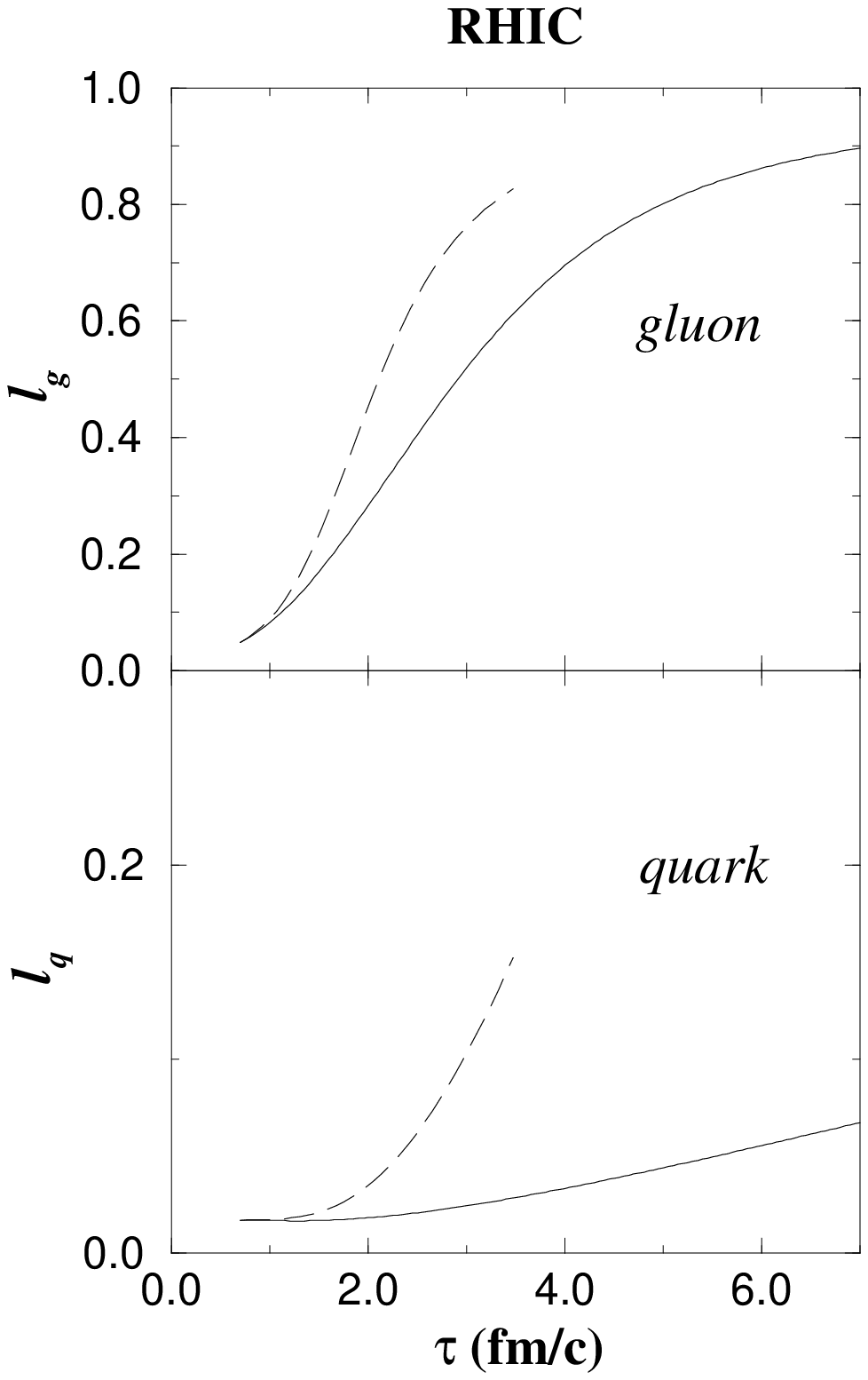,width=1.64in}
}}
\caption{Variation of the gluon and quark fugacities, $l_g$ and $l_q$,
with time. The solid (long dashed) lines are the results of fixed 
(evolving) $\a_s$.}
\label{f:fug}
\efig

As for thermalization or kinetic equilibration, there is no
one parameter as in chemical equilibration to parametrize the
degree of thermalization, so an indicator of the degree of 
thermalization must be found. Usually, this is done by examining
the slope of the particle $p_T$-spectra or that of the log
of the particle distributions. In our case, we choose instead
to plot the longitudinal to transverse pressure ratios because 
the pressure must be the same in all directions when the system 
is in kinetic equilibrium. So the closer are these ratios to 
unity, the nearer is the parton system to full thermal equilibrium. 
In \fref{f:pres}, we plot these ratios in solid lines for
gluon and quark separately. The particular shape of the curves
has to do with our rather special initial conditions, which we
take to be momentarily thermalized at the beginning.
Therefore these curves all start with a ratio of $1.0$. As the
expansion sets in, the system is driven out from the initial
transient kinetically equilibrated state and the pressures are no 
longer isotropic. The system naturally responds to the expansion
and increases the collision rate. At some point, the latter
dominates over the expansion and the curves all approach unity
again after the dip. We are more interested in the final
approach to kinetic equilibrium. As can be seen, the gluons
are better equilibrated than the quarks, which are again
not so good. 

\bfig
\centerline{
\hbox{
\epsfig{figure=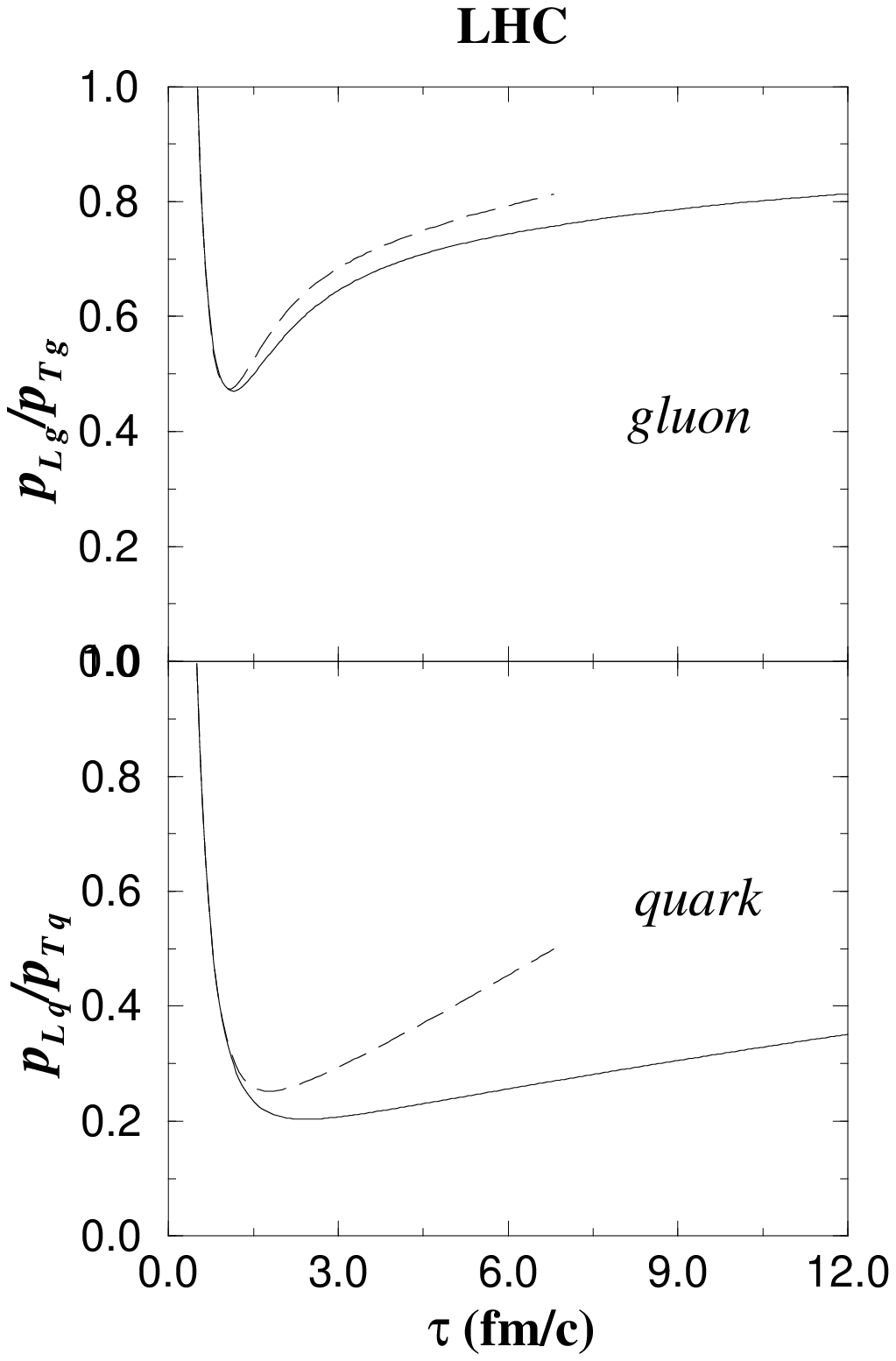,width=1.6in} \ \
\epsfig{figure=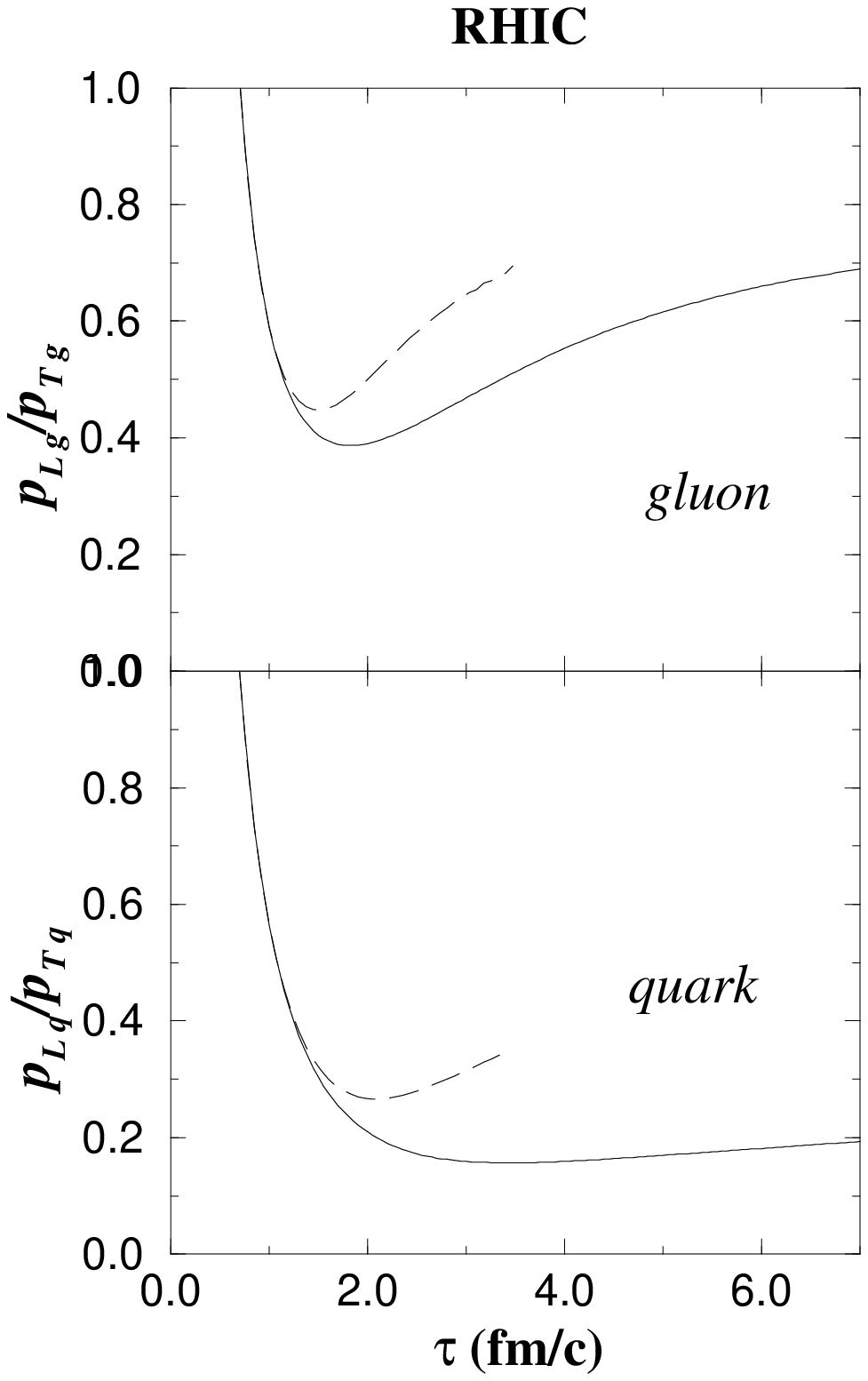,width=1.52in}
}}
\caption{Variation of the ratios of longitudinal to transverse pressure 
for gluon and quark with time. The solid (long dashed) lines are the 
results of fixed (evolving) $\a_s$.}
\label{f:pres}
\efig

So after we used some reasonable initial conditions, an
essentially perturbative time evolution scheme and an appropriate
value of the coupling, the state of equilibrium of the
parton phase is reasonably good for gluons but not so good for
the quarks. One wonders whether it is possible to do better
than this. Evidently, the closer is the plasma to full equilbrium
the better. In fact, the answer is yes. Improvement is possible 
because what we have done so far, although apparently reasonable,
is not entirely correct for a time-evolving system. It is this
last aspect, which is unusual within the framework of
perturbative calculations of strong interactions, that one could
easily overlook \cite{biro,shury,wong1,wong2}
and indeed has been the case until recently \cite{wong3}. 

In fact, during the time evolution of the system that produced
the previous results, the average parton energies dropped
significantly by at least $1.0$ GeV and so the system
underwent substantial changes. It is therefore very unlikely
that the average momentum transfer in an average parton
collision can stay fixed at around $2.0$ GeV throughout.
If so, a way must be derived to replace the fixed value of
$\a_s=0.3$ used to obtain the previous results. A simple
recipe naturally suggests itself, which is to take the
average momentum transfer $\langle Q \rangle$ to be
given by the average parton energy $\langle E \rangle$.
Then one substitutes this into the one-loop running coupling 
expression
\be \a_s (\langle E \rangle) = 
    {{4\p} \over {\b_0 \ln (\langle E \rangle^2/\Lambda_{QCD}^2)}}
\ee
to obtain an interaction strength entirely determined by the
system. As a consequence, this strong coupling will evolve
with the plasma and as such this new approach is more 
self-consistent.

The new results of the time evolution scheme with the 
self-consistent coupling are the long dash curves in \fref{f:fug}
and \fref{f:pres}. The dashed curves in \fref{f:fug} and those
in \fref{f:pres} during the final approach to equilibrium all
rise faster with time than when the plasma is evolved with
a fixed $\a_s$ (solid line). Not only is that the case, the
final state of equilibrium is markedly better than before
at both colliders. So a self-consistent time evolution speeds 
up and improves the equilibration of the parton plasma. There are 
other associated effects on the plasma that go along with the 
aboves but we shall not discuss them here. We refer the reader to 
\cite{wong3}.

\section{SU(3) non-Abelian parton plasma is an unique many-body system}

In the previous section, we saw how the evolving coupling improved
the equilibration of the parton plasma. This is but a manifestation
of an unique property of a many-body system governed by the
QCD Lagrangian. In the Boltzmann equation, it is the collision
terms that are responsible for bringing the plasma into equilibrium. 
The collision terms are all made up of the difference between 
the reactions going one way and in the reverse direction.
In an ordinary many-body system, as equilibrium is approached,
the forward and backward reaction rates get closer and closer
to each other and the equilibration rate will become slower
and slower as a result. On the contrary in a parton plasma,
because the interactions are mediated via SU(3) non-Abelian
gauge bosons, the interaction strength varies with the scale
of the interactions. The collision terms of such a system
are therefore each given by a certain power of a varying
$\a_s$ multiplied by the difference of the rate of the 
forward and backward reaction. As equilibrium
is approached, the reaction rates in both directions tend
to equalize, however, the equilibration rate of a parton
plasma does not slow down, unlike an ordinary plasma,
because the increasing strength of the interactions
caused by the lowering of the average energy of the system
is able to more than compensate for the otherwise equalization
of the forward and backward reactions or in other words, 
the slowing down of the equalibration rate. The manifestation
of this effect can be most easily seen on the variation of
the collision time with time (see \cite{wong3}).
Therefore the parton plasma exhibits a most unusual
feature of the tendency to equilibrate faster and faster.
Obviously, this cannot continue forever and at some
point the equilibration of the quarks and gluons will
be interrupted by the start of the deconfinement 
phase transition.

\section{Out-of-equilibrium effect on photon production}

\bfig
\centerline{
\hbox{
\epsfig{figure=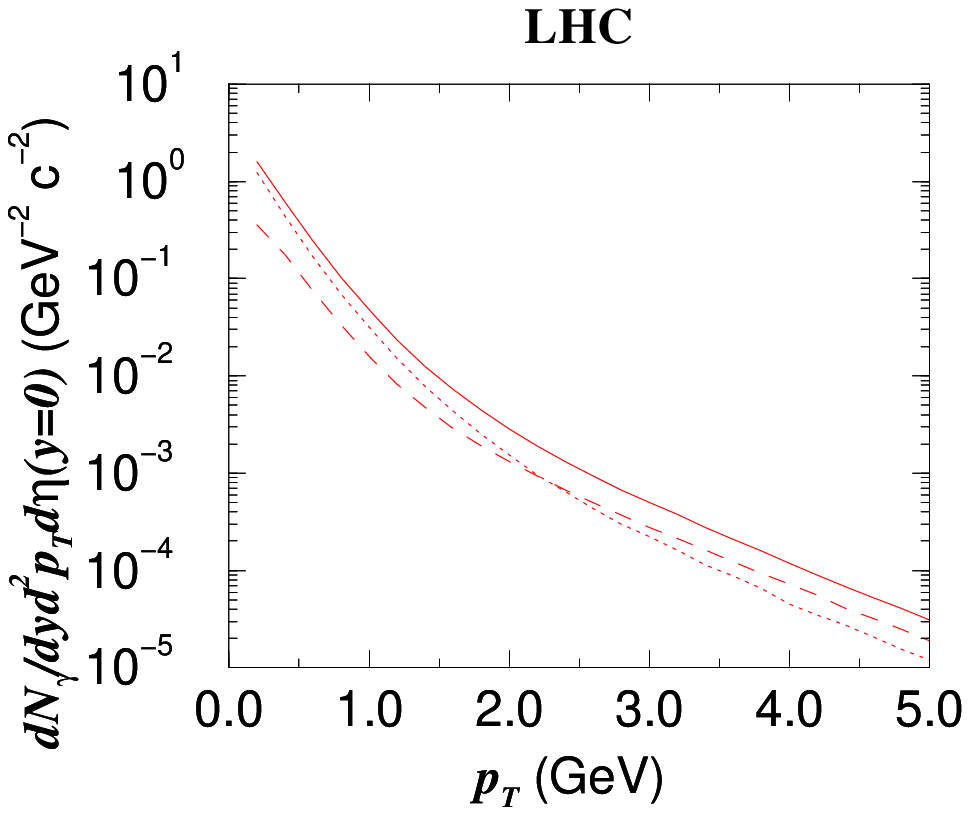,width=1.55in} \ \
\epsfig{figure=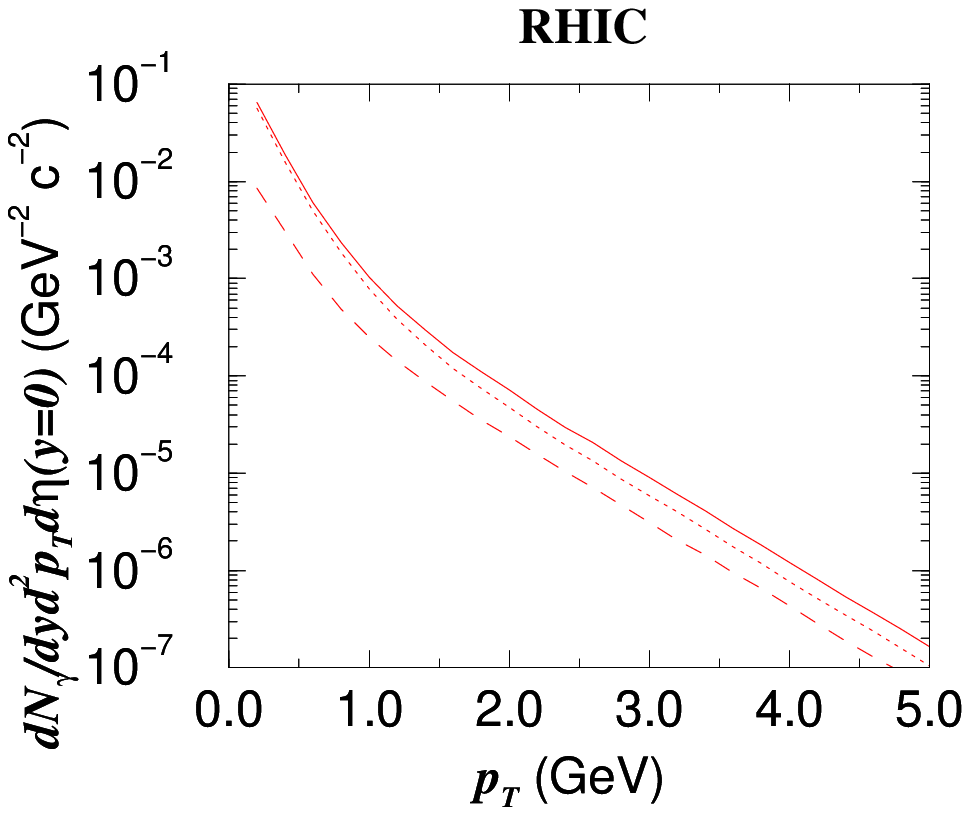,width=1.55in}
}}
\caption{Photon production at LHC and at RHIC from a parton
plasma not in equilibrium. The dotted (dashed) lines are from
Compton scattering (annihilation) contribution. The solid lines
on the top are the total contributions. The contribution from 
Compton scattering does not dominate over that from annihilation
at LHC.}
\label{f:photon}
\efig

We have shown the time evolution of a parton plasma at the
two colliders. All the information of the system
is contained in the particle distributions and from which
we learned about the plasma's state of equilibrium.
With the knowledge of the momentum distributions of
the particles, one can work out the rates of particle production
such as photons, dileptons etc..
Also our method of evolving the plasma in time, imposed
no equilibrium requirement on any aspects of equilibration
and hence the form of the distributions. As such, a direct
comparison of particle productions of a plasma that is out
of equilbrium with an equilibrated one can be done.
We will show an out-of-equilibrium effect on photon 
production and explain its importance. 

Photon production from the plasma is dominantly through
Compton scattering and annihilation contribution. The
total production rate from these two contributions is 
given by 
\bea E {{d^7 N} \over {d^3 p d^4 x}} \fx &=& \fx \frac{1}{2(2\p)^3}
     \int \intkso \intkst \intksth                          
     (2\p)^4 \d^{(4)} (k_1+k_2-k_3-p) \nonum
     \fx & \times & \fx 
   \bigg \{ 2 f_g(k_1,\t) f_q(k_2,\t) (1-f_q(k_3,\t))       
     |\cm_{gq\lra q \g}|^2        \nonum
     \fx & & \fx \;  
          + f_q(k_1,\t) f_{\bar q}(k_2,\t) (1+f_g(k_3,\t))  
     |\cm_{q\bar q\lra g \g}|^2                
   \bigg \}  \; .
\label{eq:photon}
\eea

In ref. \cite{mull}, a kinetically equalibrated plasma was 
studied and the photon rate calculated. It was shown in their 
Fig. 8 and 9 that the contribution from Compton scattering
was far more important than the annihilation contribution
both at LHC and at RHIC energies. If one now compares 
these figures with the one we have here in \fref{f:photon},
one can immediately see that in our plots, Compton scattering
contribution does not dominate over that of annihilation.
In fact at higher $p_T$ at LHC, annihilation contribution
is the more important. At RHIC in the same $p_T$ range,
although Compton scattering is larger, it is not 
by the same large amount as shown in ref. \cite{mull}. 
The reason for these differences between our photon
production and those in ref. \cite{mull} is because our
plasma is not in kinetic equilibrium but that in 
ref. \cite{mull} was.  We will explain this in the next
paragraph. 

In \eref{eq:photon}, the Compton and annihilation
contribution to photon yield differ by various factors
such as the interaction matrix elements, the Pauli
blocking and Bose-Einstein stimulated emission, and
one of the initial distributions. In fact, the matrix
elements do not differ significantly numerically
and the Pauli blocking and Bose-Einstein stimulated
emission also provide no great suppression or enhancement.
The source responsible for the difference seen in the
two different sets of figures comes actually from the
gluon distribution in Compton scattering and the
initial quark distribution in the annihilation contribution. 
To explain this more clearly, we use the observation that 
the difference in the figures is most prominent for larger
$p_T$ photons --- annihilation contribution is actually
larger than Compton scattering at LHC in \fref{f:photon}
--- so we can concentrate in this $p_T$ range. 
Emission of larger $p_T$ photons requires more energetic 
incoming partons, this let us simplify the explanation by 
assuming Boltzmann form for the incoming particle distributions
$f \sim l \exp \{-k^0/T\}$, while for the outgoing ones,
we will replace them by unity. In reality, Pauli blocking
and Bose-Einstein stimulated emission do have some
effects but they are not so important in comparison
with the one that we are going to describe so it is best
to leave them out. Since we are comparing the two
contributions, we form the ratio
\be {\mbox{Annihilation} \over \mbox{Compton}}
   \sim \frac{l_q}{l_g} \exp \; \{k^0 (1/T_g-1/T_q)\}
\label{eq:ratio}
\ee
after taking $|\cm_{gq \lra q\g}| \sim |\cm_{q\bar q\lra g\g}|$.
Because of the initial gluon dominance and the stronger interaction
amongst gluons due to colour, $l_q/l_g << 1$ always and 
therefore if the exponential factor is not present, Compton 
scattering indeed dominates over annihilation contribution.
In a kinetically equilibrated parton plasma, the system has only 
one temperature $T_g = T_q$, so the exponential factor is unity
and we get the results shown in Fig.~8 and 9 of ref. \cite{mull}. 
In our present case, the plasma is not in kinetic equilibrium
but we can most simply consider them as a mixed fluid of gluons
and quark-antiquarks at different temperatures.
The gluon temperature is higher initially but will eventually
be lower than that of the quarks because of gluon multiplication
and their conversion into quark-antiquark pairs. When that 
happens, for a photon with a high enough $p_T$, the incoming 
parton energies will be sufficiently large that the product in
\eref{eq:ratio} will be larger than unity and the photon
rate from annihilation will be larger than that from
Compton scattering. This is what one sees in our result
in \fref{f:photon} at LHC. In the same figure at RHIC,
there is not sufficient time for the two temperatures to be
in the correct range from each other so that the exponential 
factor is able to compensate for the fugacity ratio. The result 
is Compton scattering remains larger than the annihilation
contribution but it is definitely by a lesser amount than that
shown in Fig. 8 and 9 of ref. \cite{mull}.

We see that the out-of-equilibrium effect tends to enhance the 
annihilation contribution at higher $p_T$ and hence the total 
photon yield in that $p_T$ range. Therefore photon 
production of partonic origin should have a better chance
to compete with the hard photons from the initial collisions
and those fragmented off minijets. The window for observing 
photons from deconfined matter will be widened as a result.

\section*{Acknowledgments}
The author would like to thank the organizers for a very 
enjoyable symposium and for everyone, especially C. Ktorides and 
C.N. Papanicolas, and others too numerous to mention individually 
at the Nuclear and Particle Physics Section of the University of Athens 
and at the Institute IASA for kind hospitality during his stay there.

\end{document}